\documentclass[%
 reprint,
 amsmath,amssymb,
 aps,
]{revtex4-2}

\usepackage{graphicx}% Include figure files
\usepackage{dcolumn}% Align table columns on decimal point
\usepackage{bm}% bold math
\usepackage{hyperref}

\begin{document}

\preprint{APS/123-QED}

\title{Steady-state squeezing transfer in hybrid optomechanics}% Force line breaks with \\

\author{Hugo Molinares}%
\email{hugo.molinares@mayor.cl}
\affiliation{%
Centro de Optica e Información Cuántica, Universidad Mayor,\\
camino la Piramide 5750, Huechuraba, Santiago, Chile
}%
\author{Vitalie Eremeev}%
\email{vitalie.eremeev@udp.cl}
\affiliation{%
Facultad de Ingeniería y Ciencias, Universidad Diego Portales, Av. Ejercito 441, Santiago, Chile
}%
\author{Miguel Orszag}%
\email{miguel.orszag@umayor.cl}
\affiliation{%
Centro de Optica e Información Cuántica, Universidad Mayor,\\
camino la Piramide 5750, Huechuraba, Santiago, Chile
}%
\affiliation{%
Instituto de Física, Pontificia Universidad Católica de Chile, Casilla 306, Santiago, Chile
}
\date{\today}

\begin{abstract}
A hybrid scheme is presented that allows the transfer of squeezed states (TSS) from the mechanical part to an optical cavity in the steady-state. In a standard optomechanical scheme, a three-level atom acts as an intermediate element for TSS. Two different procedures are developed that allow the visualization of the TSS effect: In the first one, we apply a coherent pump of squeezed phonons in our hybrid system, and the second method is achieved by placing the system in contact with a phonon squeezed bath. Our model and procedures show that in optomechanical systems TSS can be achieved with a high fidelity. 
\end{abstract}

\maketitle

%\tableofcontents

%\section{\label{sec:level1}INTRODUCTION}

With the remarkable state-of-the-art of the hybrid systems \cite{RevModPhys.86.1391,Mun2018,Mir2020} composed by mechanical oscillators (MO), cavities, spins, etc., it is becoming more and more feasible to control such systems in their quantum regime in search for non-classical features of their elements. Besides MO have the ability to easily interact with a wide range of physical systems, such as ultracold atomic BECs \cite{Hun2010}, superconducting qubits \cite{Mir2020}, spin states in quantum dots \cite{Ste2009, Car2018} and color/NV centers \cite{MacQ2015, Adi2017, Mun2018}, cavity fields in the opto-mechanical systems \cite{Mar2018, Agg2020}, etc. %Actually, the cooling down to sub-Kelvin temperatures of relatively large mechanical objects can be reached through different technics as feedback cooling \cite{Li2011, Guo2019}, by a dynamical decoupling in spin-mechanics \cite{MacQ2015, Rao2016}, so allowing the mechanics to oscillate with very low number of quanta excitations for different applications in the framework of the hybrid devices. The aforementioned hybrid mechanical architectures are acquiring an increasing attention in order to understand the foundations of quantum theory, such as macroscopic quantum superpositions \cite{Nak1999, Liao2016, Mon2017} and quantum correlations at macroscopic scales \cite{Sca2013, Mar2018} – a very well-known questions since Schroedinger's times on quantum mechanics. 

The hybrid setups are of paramount importance regarding the development of the quantum technologies. Therefore, to move forward the limits of the fields such as quantum computing, quantum networks, quantum metrology and sensing, a very demanded tool should be the high-fidelity transfer of the quantum states between the components of a hybrid system. For instance, in the realm of quantum networks, the mechanical object can serve as light-matter transducer \cite{Sta2010, Rie2016} or to map/encode information from a qubit \cite{Ree2017, Mir2020}. Particularly, spin-mechanical systems %(a spin like qubit coupled to a MO) 
have attained major attention, mainly because spin systems exhibit long-coherence times and they can be easily manipulated and read-out \cite{Rob2011, Rao2016}. 
A rapidly emerging field is the quantum metrology and sensing \cite{Gio2006} by using the quantum states and protocols with the aim to overcome the precisions unreachable by the classical sensing. There are several experiments proving the sensing near and beyond the standard quantum limit, e.g. \cite{And2017, Kam2017, Mas2019}. % Nevertheless, the quantum metrology needs more elaborated techniques and methods in order to improve the precisions from the shot-noise to the Heisenberg limit. 
Commonly the quantum sensors are based on the photonics setups, involving cavities, atoms, photons, trapped ions, and solid-state systems with electrons, superconducting junctions, etc. On the other hand, the hybrid systems like opto/spin-mechanical, opto-electromechanical setups, etc., actually become more attractive for their effectiveness and usefulness for a wide range of quantum applications, from gravitational wave detectors \cite{Ligo2013, Gro2013} to force microscopes \cite{All2018, He2020}, hence considered as the leader candidates for the quantum metrology and sensing. 

In this context, the squeezing of the modes in a hybrid system, and particularly the squeezing transfer between them, is of a major importance and applicability. On the one hand, the preparation of the mechanical and light modes in the squeezed states, has been widely investigated theoretically \cite{Lu2015, Liu2018, Zhang2020, Wang2020} and is nowadays experimentally feasible in versatile hybrid setups \cite{Bro2012, Saf2013, Pur2013, Pir2015, Woll2015, Ock2018, Agg2020}. On the other hand, there are less proposals for squeezing transfer between different elements in a hybrid system, e.g. some theoretical work over the last two decades \cite{Mas2001, Ran2005, Wal2010, Mom2019, Bai2019}. To advance in this direction, in the present work we propose an optomechanical protocol of high-fidelity squeezing transfer from the mechanical mode to the cavity field. Our results show that such a transfer could be realized by a coherent mechanical squeezed pump, as well as under the coupling with a squeezed phononic bath.   

\begin{figure}[t]
\includegraphics[width=1\linewidth]{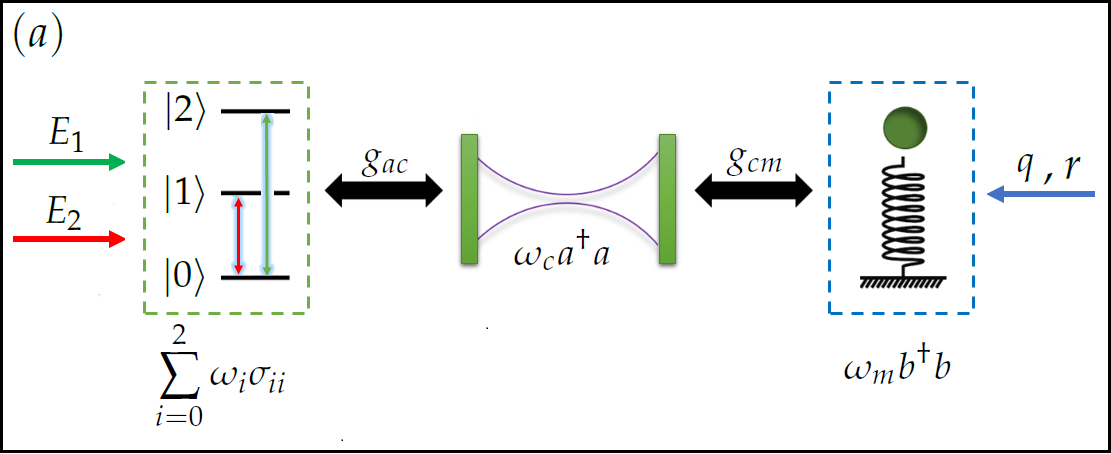}

\includegraphics[width=1\linewidth]{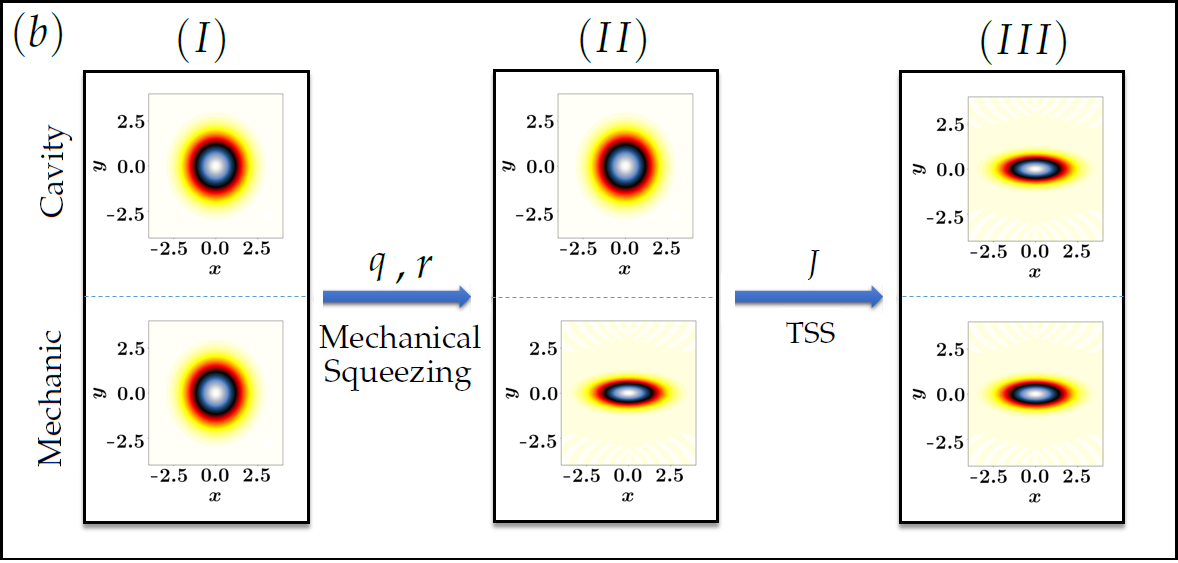}
\caption{$(a)$ Schematic diagram of a hybrid cavity-atom-mechanics system. $(b)$ Visualization of Wigner function to
explain how squeezing is transferred from the mechanical part to the cavity field. $(I)$ Initial states of the cavity field and mechanical part. $(II)$ States in presence of the mechanical squeezing. The mechanical part of the system is squeezed via $i)$ Applying a coherent squeezed pump of the form $\mathcal{H}_{q}$, and $ii)$ by coupling the system to an incoherent pump, that is, a squeezed mechanical reservoir. $(III)$ Final steady states. The squeezed state is transferred from the mechanical part to the cavity field through the tripartite atom-photon-phonon interaction coupling strength $J$.}
\label{fig1}
\end{figure}

As illustrated in Fig. \ref{fig1}$(a)$, we consider a hybrid system where a confined photon mode is coupled both to two upper levels of the three-level atom and to a mechanical oscillator as ($\hbar=1$)
\begin{eqnarray}\label{base}
    \mathcal{H}_{0}&=&\sum_{i=0}^{2}\omega_{i}\sigma_{ii}+\omega_{c}a^{\dagger}a+\omega_{m}b^{\dagger}b \nonumber\\
    &&+\imath g_{ac}\left(a\sigma^{+}_{21}-a^{\dagger}\sigma^{-}_{21}\right)
    -\imath g_{cm}a^{\dagger}a\left(b^{\dagger}-b\right),
\end{eqnarray}
where $\sigma^{\pm}_{ij}$ are the ladder operators for the atom, $a(b)$ is the annihilation operator of the cavity (mechanical) mode of frequency $\omega_{c}(\omega_{m})$ and $\omega_{i}$ are the energy levels of the three-level atom. $g_{ac}$ and $g_{cm}$ are the atom-cavity and optomechanical coupling strengths, respectively.

When this hybrid system is driven by the coherent pump $\mathcal{H}_{q}=\imath q (b^{\dagger2}-b^{2})$ ($q$ is proportional to the driving field strength), the mechanical resonator is prepared in a squeezed state \cite{Walls}. The full Hamiltonian in the interaction picture (rotating at the mechanical frequency $\omega_{m}$) takes the form

\begin{eqnarray}\label{h1}
    \mathcal{\tilde{H}}_{1}&=&\imath g_{ac}\sigma_{21}^{+}ae^{\imath\Delta t}e^{-\imath F^{*}(t)}
    +\imath q b^{\dagger 2}+2\imath Ka^{\dagger}ab^{\dagger}\eta\nonumber\\
    && + H.c.
\end{eqnarray}
Here we used the Hermitian operator $F\equiv\frac{g_{cm}}{\omega_{m}}\left(b^{\dagger}\eta^{*}+b\eta\right)$, with $\eta\equiv e^{\imath\omega_{m}t}-1$; $\Delta\equiv\omega_{2}-\omega_{1}-\omega_{c}$ is the detuning, and the parameter $K\equiv q\cdot g_{cm}/w_{m}$ depends on the optomechanical coupling and the amplitude of the coherent squeezing pump. Considering the experiments in optomechanics \cite{RevModPhys.86.1391}, it is expected that $K$ takes very small values.

In the following, we assume that the optomechanical coupling $g_{cm}$ is much smaller than the mechanical frequency $\omega_{m}$, so that
$e^{-\imath F^{*}(t)}\approx1-\imath\frac{g_{cm}}{\omega_{m}}\left(b^{\dagger}\eta^{*}+b\eta\right)$. Using the rotating wave approximation under the blue-detuned regime $\Delta=\omega_{m}$, and keeping only the time independent terms, the above Hamiltonian becomes
%\begin{equation}\label{h2}
%    \mathcal{\tilde{H}}_{2}=\imath J\left(\sigma_{21}^{+}ab^{\dagger}-\sigma_{21}^{-}a^{\dagger}b\right)+\imath q \left(b^{\dagger 2}-b^{2}\right)-2\imath Ka^{\dagger}a\left(b^{\dagger}-b\right),
%\end{equation}
\begin{equation}\label{h2}
    \mathcal{\tilde{H}}_{2}=\imath J\sigma_{21}^{+}ab^{\dagger}+\imath q b^{\dagger 2}-2\imath Ka^{\dagger}ab^{\dagger} + H.c.,
\end{equation}
where $J\equiv g_{ac}\cdot g_{cm}/w_{m}$ is the tripartite atom-photon-phonon interaction strength.

In order to produce squeezed light in the optical cavity, two lasers $E_{1}$ and $E_{2}$ (proportional to the field strengths) are introduced in the system, which are resonant with the transitions of the levels $|2\rangle \longleftrightarrow |0\rangle$ and $|1\rangle \longleftrightarrow |0\rangle$, respectively. These coherent drives are described by the Hamiltonian in the interaction picture $\mathcal{\tilde{H}}_{E}=\imath E_{1}\left(\sigma_{20}^{-}-\sigma_{20}^{+}\right)+\imath E_{2}\left(\sigma_{10}^{-}-\sigma_{10}^{+}\right)$. If we now include the dissipation caused by the system-environment coupling, the dissipative dynamics of the hybrid quantum system is described by the Markovian master equation 
\begin{eqnarray}\label{dinamica} 
    \frac{d\rho}{dt}&=&-\imath[\mathcal{\tilde{H}}_{2}+\mathcal{\tilde{H}}_{E},\rho]+\frac{\gamma_{21}}{2}\mathcal{L}[\sigma_{21}]\rho\nonumber\\
    &&
    +\frac{\gamma_{10}}{2}\mathcal{L}[\sigma_{10}]\rho+\frac{\kappa_{a}}{2}\mathcal{L}[a]\rho+\frac{\kappa_{b}}{2}\mathcal{L}[b]\rho,
\end{eqnarray}
where $\rho$ is the density matrix of the hybrid system and the common Lindblad dissipative terms are defined by: $\forall\mathcal{O},$ $\mathcal{L}[\mathcal{O}]=2\mathcal{O}\rho \mathcal{O}^{\dagger}-\mathcal{O}^{\dagger}\mathcal{O}\rho-\rho \mathcal{O}^{\dagger}\mathcal{O}$, with all the baths at $n_{th}=0$. Here $\gamma_{21}$ $(\gamma_{10})$ correspond to spontaneous emission rates from level 2 to 1  (level 1 to 0), respectively and $\kappa_{a}$ $(\kappa_{b})$ are the decay rates of the optical (mechanical) mode. The approach of $n_{th}=0$ is realistic in recent experiments, for example for a hybrid system with atoms, cavity and mechanics in the regime of microwave frequencies, as in recent experiments \cite{Mir2020,Rie2016} the mechanical mode has $\omega_m\approx 30$ GHz, and by cooling the system to the temperatures of $\sim 30$ mK, one gets $n_{th}\equiv[\exp(\hbar\omega_m/k_B T)-1]^{-1}\approx 10^{-4}$.  

After building the theoretical model, we discuss the degree of squeezing present in the states of the cavity and the mechanical part. For this, we rely on numerical methods according to \cite{qutip} to solve Eq. \ref{dinamica} in the steady-state, i.e. $\Dot{\rho}=0$, and so calculating the quadrature fluctuations. Here, we study the parameter space in order to minimize fluctuations and achieve an optimal result for TSS in the stable region. 

For a better understanding the TSS effect and stability of the hybrid system, we get a set of first order differential equations from Eq. \ref{dinamica}:
\begin{align}\label{dif}
    \frac{d\langle a^{\dagger}a\rangle}{dt}&= -2J\langle a^{\dagger}b\rangle\langle\sigma^{-}_{21}\rangle-\kappa_{a}\langle a^{\dagger}a\rangle,\\
     \frac{d\langle b^{\dagger}b\rangle}{dt}&= 2J\langle a^{\dagger}b\rangle\langle\sigma^{-}_{21}\rangle
     + 4q\langle b^{2}\rangle- \kappa_{b}\langle b^{\dagger}b\rangle,\\
     \frac{d\langle a^{2}\rangle}{dt}&=  -2J\langle ab\rangle\langle\sigma^{-}_{21}\rangle-\kappa_{a}\langle a^{2}\rangle,\\
     \frac{d\langle b^{2}\rangle}{dt}&= 2J\langle ab\rangle\langle\sigma_{21}^{+}\rangle+4q\langle b^{\dagger}b\rangle-\kappa_{b}\langle b^{2}\rangle+2q,
\end{align}
\begin{align}
     \frac{d\langle a^{\dagger}b\rangle}{dt}&= 
      -J\left(\langle b^{\dagger}b\rangle-\langle a^{\dagger}a\rangle\right)\langle\sigma^{+}_{21}\rangle+2q\langle ab\rangle\nonumber\\
      &-\frac{\kappa_{a}}{2}\langle a^{\dagger}b\rangle-\frac{\kappa_{b}}{2}\langle a^{\dagger}b\rangle,\\
      \frac{d\langle ab\rangle}{dt}&= -J\left(\langle b^{2}\rangle\langle\sigma^{-}_{21}\rangle-\langle a^{2}\rangle\langle\sigma^{+}_{21}\rangle\right)+2q\langle a^{\dagger}b\rangle\nonumber\\
      &-\frac{\kappa_{a}}{2}\langle ab\rangle-\frac{\kappa_{b}}{2}\langle ab\rangle,
      \label{dif2}
\end{align}

where we have considered the factorization of the form $\langle a^{\dagger}b\sigma^{-}_{21}\rangle=\langle a^{\dagger}b\rangle\langle\sigma^{-}_{21}\rangle$ (adiabatic approximation) and since $\omega_{m}\gg \left\{g_{cm},q\right\}$ we have considered $K=0$. 
%%%%%%%%%%%Figure 2%%%%%%%%%%%%%%
\begin{figure}[t]
\centering
\includegraphics[width=0.48\linewidth]{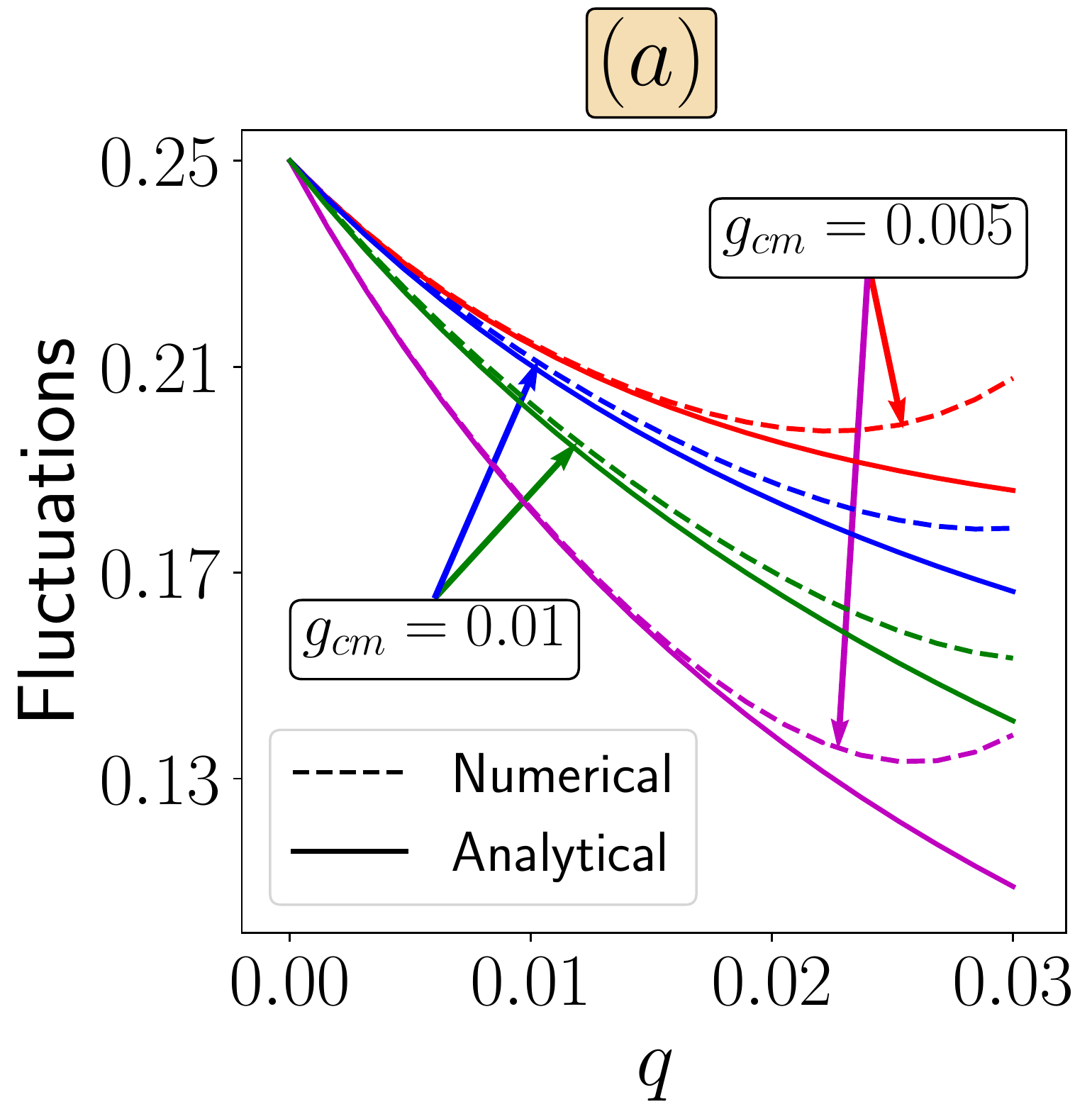}
\includegraphics[width=0.48\linewidth]{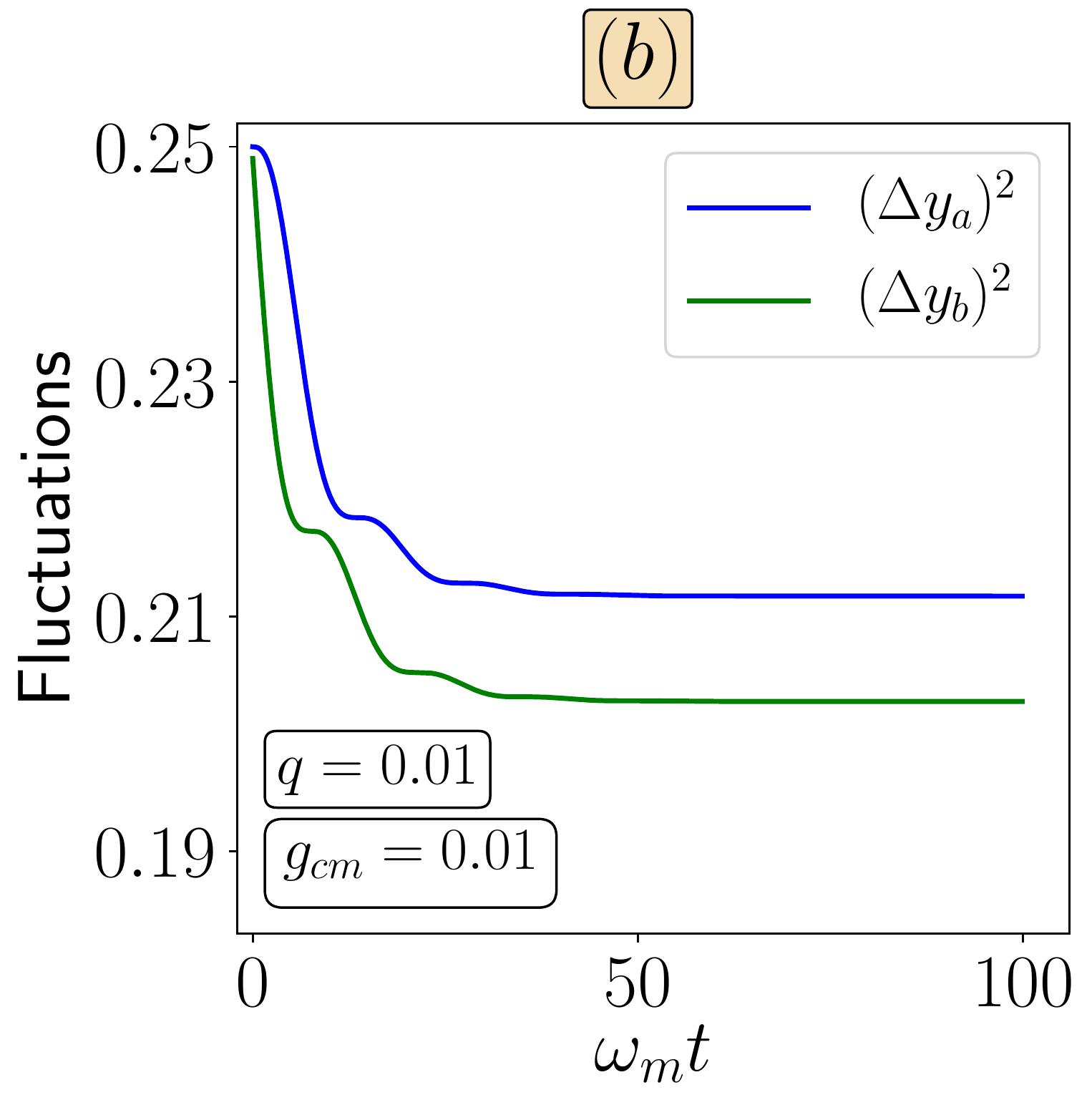}
\caption{$(a)$ Quadrature fluctuations as functions of the parameter $q$, for the cavity $\left(\Delta y_{a}\right)^{2}$ (red and blue curves) and for the mechanics $\left(\Delta y_{b}\right)^{2}$ (green and magenta curves). $(b)$ Time evolution of the quadrature fluctuations. All the parameters are scaled by $\omega_m$: $g_{ac}=10^{2}$, $\kappa_{a}=0.2$, $\kappa_{b}=\kappa_{a}\cdot 10^{-2}$, $\gamma_{10}=20$, $\gamma_{21}=0$, $E_{1}=E_{2}=25$.}
\label{fig2}
\end{figure}
%%%%%%%%%%%%%%%%%%%%%%%%%%%%%%%%%%

By considering $\left\{E_{1}, E_{2}, \gamma_{10}\right\}\gg \left\{J, q, \kappa_{a}, \kappa_{b}\right\}$ and $\gamma_{21}=0$, we get the following semiclassical results for the quadrature fluctuations in steady-state:
\begin{align}
    \left(\Delta y_{a}\right)^{2}&=\frac{m+n+s+1}{4(m+1)(n+1)},\\
    \left(\Delta y_{b}\right)^{2}&=\frac{n+s}{4(m+1)(n+1)},
\end{align}
where
\begin{align}
    m&=\frac{4q}{\kappa_{a}+\kappa_{b}},\\
    n&=\frac{\kappa_{a}\kappa_{b}+4J^{2}\langle\sigma_{21}^{+}\rangle^{2}}{4q\kappa_{a}},\\
    s&=\frac{\kappa_{b}}{\kappa_{a}+\kappa_{b}},\\
    \langle\sigma_{21}^{+}\rangle&=\frac{2E_{1}E_{2}}{\gamma_{10}^{2}+4\left(E_{1}^{2}+E_{2}^{2}\right)}.
\end{align}
The stability of the system is obtained by solving the system of equations of the form
$\frac{d\vec{x}}{dt}=M\vec{x}+c$,
where we have defined a vector $\vec{x}$ with the moments shown in the left-side of Eqs. \ref{dif}-\ref{dif2}, $M$ is a matrix built with the elements on the right-side of Eqs. \ref{dif}-\ref{dif2} and $c$ is a constant vector arising from the constant term $2q$ in the equation
for $\langle b^{2}\rangle$. The steady-state is only achieved if the real part of the eigenvalues of the matrix are all negative. Under the present approximations, we get a stationary state only if $4q<\kappa_{a}+\kappa_{b}$.
%%%%%%%%%%%Figure 3%%%%%%%%%%%%%%
\begin{figure}[t]
\centering
\includegraphics[width=0.486\linewidth]{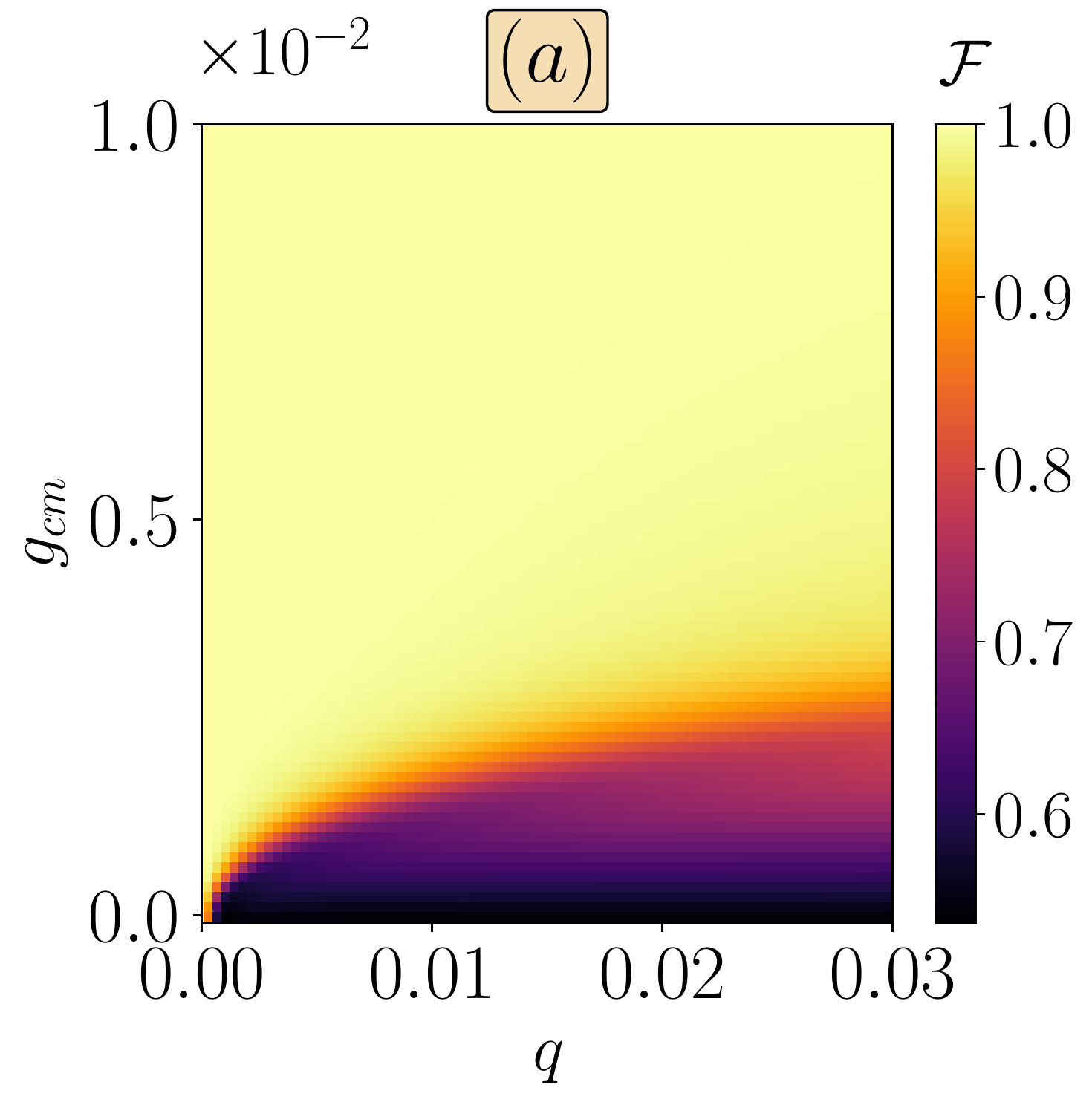}
\includegraphics[width=0.486\linewidth]{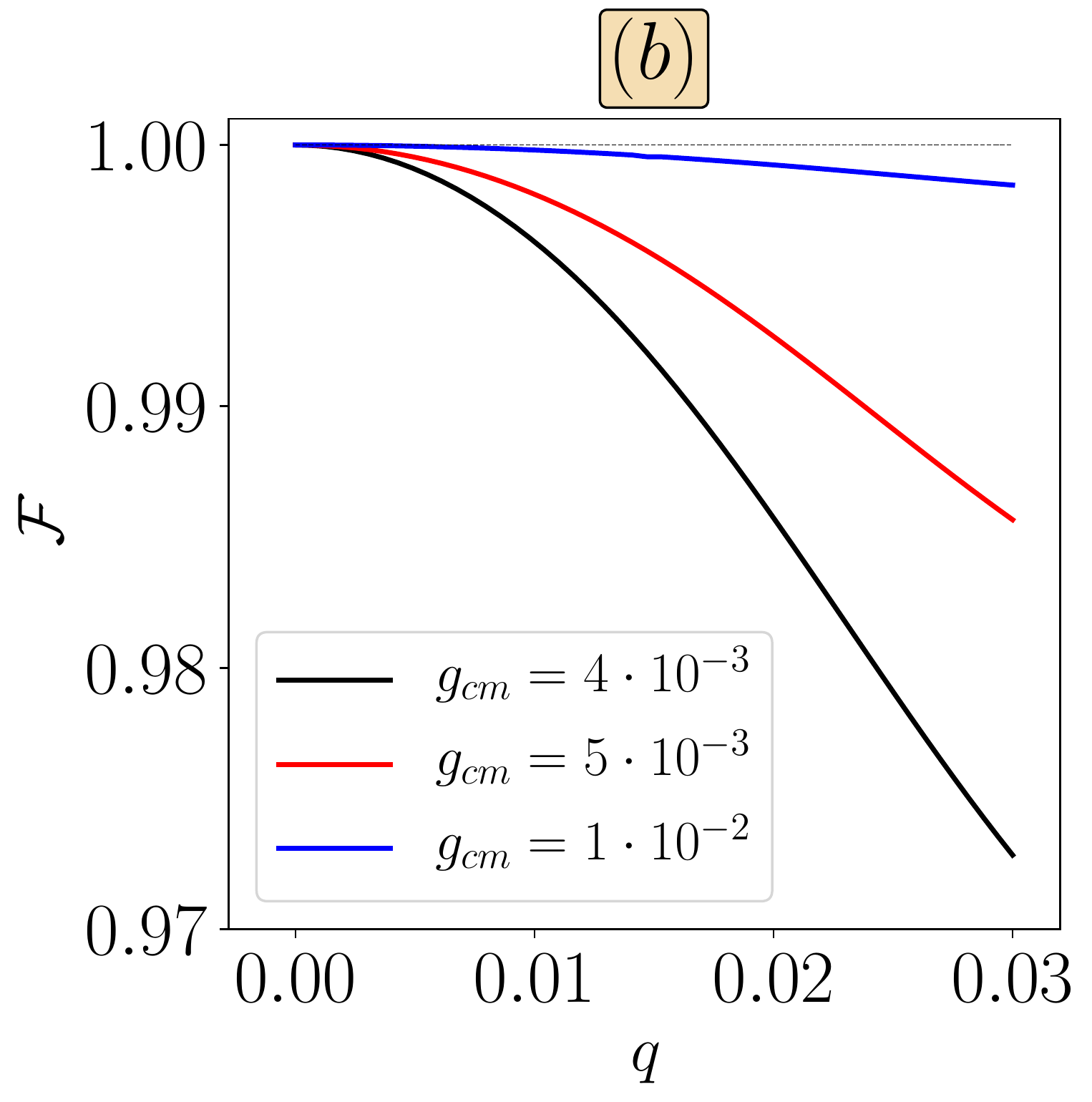}
\caption{$(a)$ Fidelity between the cavity field and mechanical state as a function of the optomechanical coupling, $g_{cm}$, and squeezing pump, $q$. As seen in $(b)$, an election of higher optomechanical coupling, $g_{cm}=10^{-2}$, gives the optimal TSS.}
\label{fig3}
\end{figure}
%%%%%%%%%%%%%%%%%%%%%%%%%%%%%%%%%%%%%

In panel $(a)$ of the Fig. \ref{fig2} we show the fluctuations as a function of the parameter $q$ for different couplings $g_{cm}$. One sees that for $g_{cm}=0.01$, the TSS shows an optimal performance, since the mechanical and optical fluctuations are close as well as the semiclassical solutions. In fact, the numerical solution could reach the analytical solution for a large enough Hilbert space, however, such dynamics for larger $q$ values increases dramatically the computation resources. To support the above, in panel $(b)$ of the Fig. \ref{fig2}, we show the time evolution of the quadrature fluctuations. Without loss of generality, we have fixed the parameters $q=0.01$ and $g_{cm}=0.01$. We can see how the injection of mechanical squeezing in the system modifies the original cavity and mechanical wave functions, bringing these to squeezed steady-states. Furthermore, one observes that both fluctuations are quite close to each other.

According to this last result, we investigate numerically the steady-state TSS by computing the fidelity \cite{Nielsen}, defined as $\mathcal{F}(\rho_{c}^{ss},\rho_{m}^{ss})\equiv Tr\sqrt{\sqrt{\rho_{c}^{ss}}\rho_{m}^{ss}\sqrt{\rho_{c}^{ss}}}$, where $\rho_{c}^{ss}$ and $\rho_{m}^{ss}$ are the steady-state density operators of the cavity and mechanical part. In panel $(a)$ of Fig. \ref{fig3} we show the fidelity as a function of $g_{cm}$ and $q$. Notice that the fidelity is close to unity as the optomechanical coupling increases, even in the presence of dissipation in the system. In panel $(b)$ of Fig. \ref{fig3} we show the fidelity as a function of $q$ for some values of $g_{cm}$. This result allows us to conclude that in the strong coupling regime the open quantum system allows the TSS with a fidelity close to unity.

We propose another method for performing TSS under incoherent driving of phonon squeezing, that is the hybrid system with $q=0$ and in contact with a squeezed reservoir. To treat such an effect, considering $q=0$ in the Hamiltonian (\ref{h2}), one has the following interaction Hamiltonian:
\begin{equation}
    \mathcal{\tilde{H}}_{3}=\imath J\left(\sigma_{21}^{+}ab^{\dagger}-\sigma_{21}^{-}a^{\dagger}b\right).
\end{equation}
%%%%%%%%%%%Figure 4%%%%%%%%%%%%%%
\begin{figure}[t]
\centering
\includegraphics[width=0.48\linewidth]{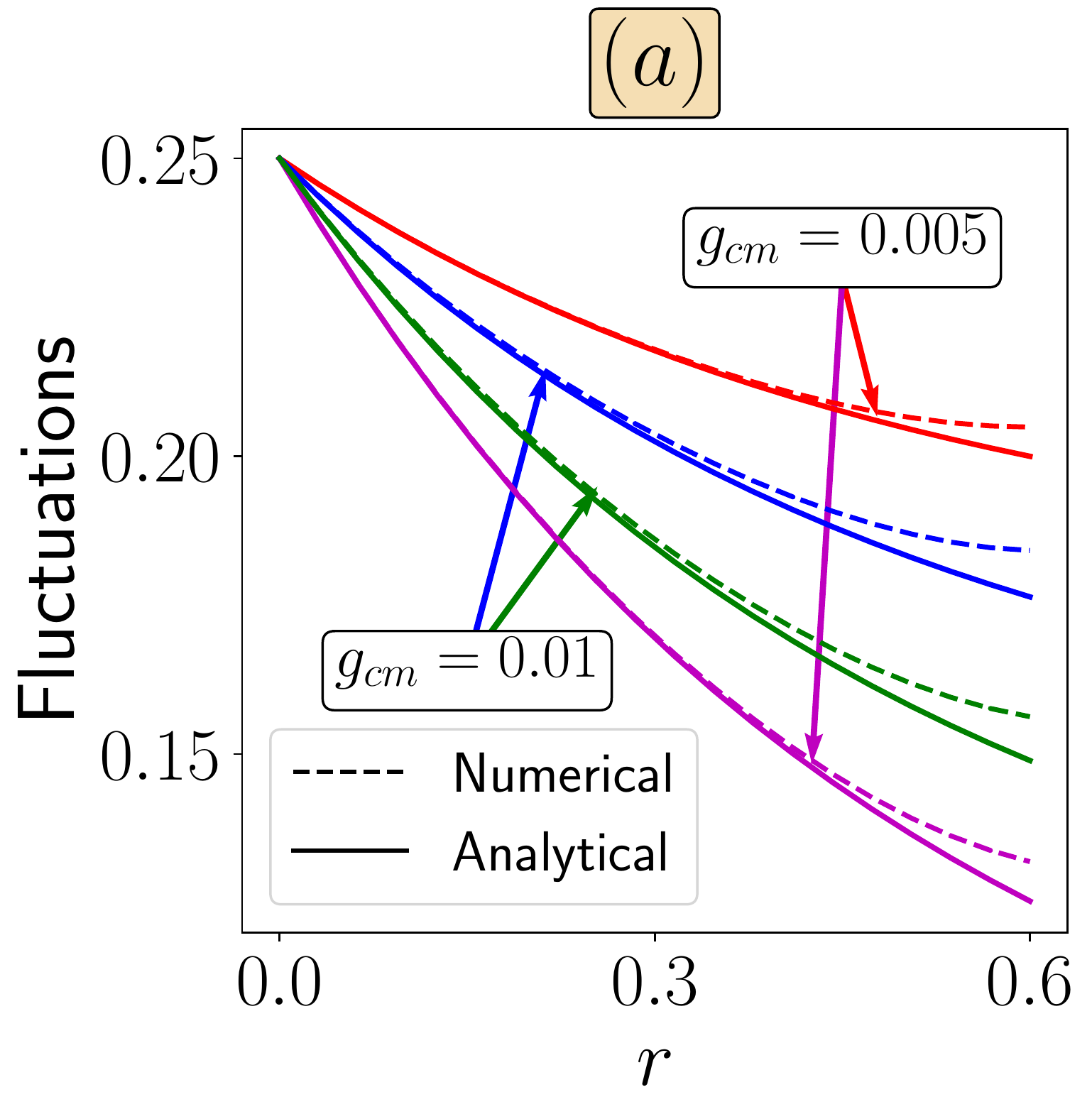}
\includegraphics[width=0.48\linewidth]{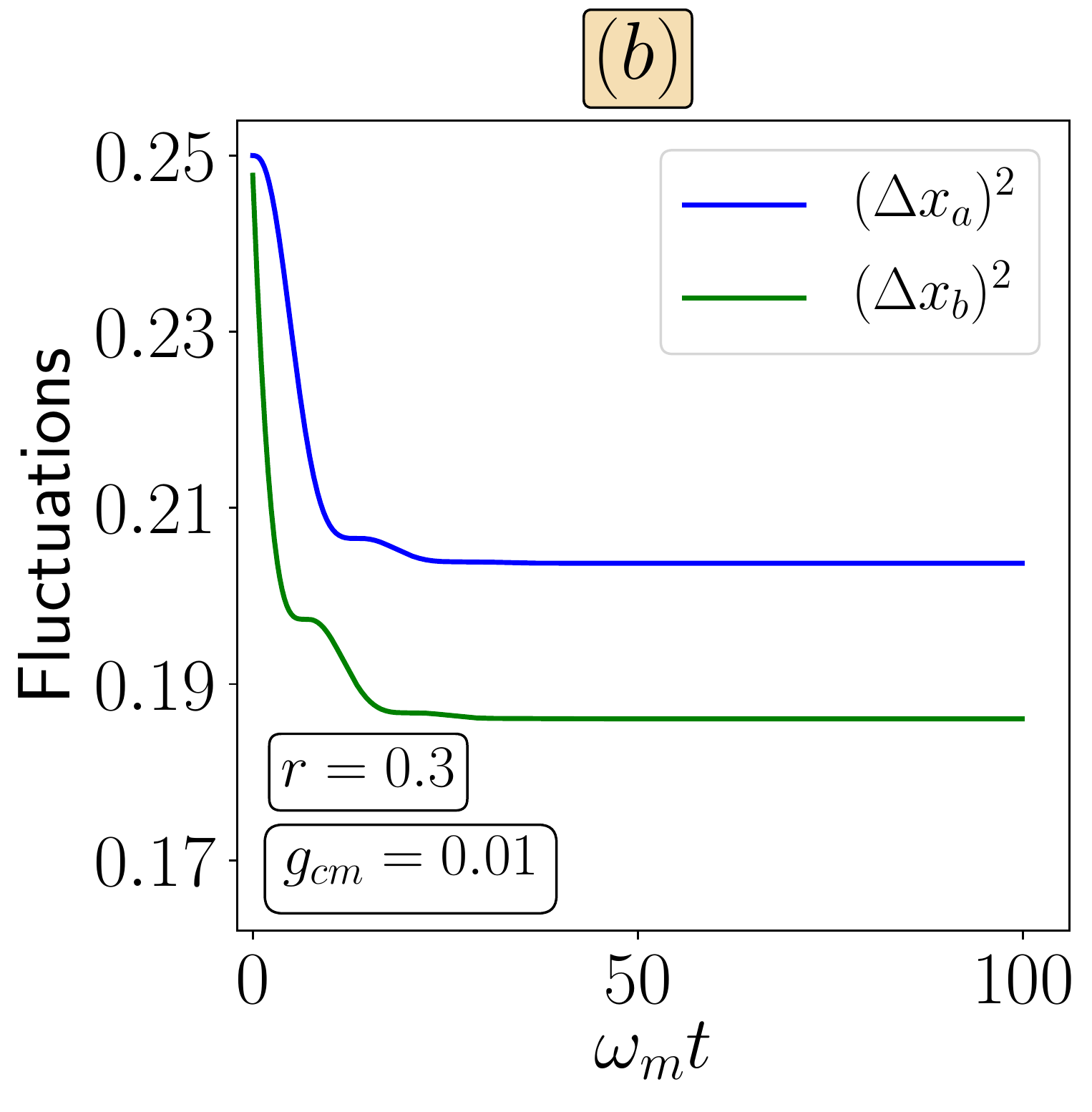}
\caption{$(a)$ Quadrature fluctuations as functions of the parameter $r$. Here $\left(\Delta x_{a}\right)^{2}$ (red and blue curves) and $\left(\Delta x_{b}\right)^{2}$ (green and magenta curves). $(b)$ Time evolution of the fluctuations evidencing the steady-state. The parameters are: $g_{ac}=10^{2}$, $\kappa_{a}=\kappa_{b}=0.2$, $\gamma_{10}=20$, $\gamma_{21}=0$, $E_{1}=E_{2}=25$, $\theta=\pi$.}
\label{fig4}
\end{figure}
%%%%%%%%%%%%%%%%%%%%%%%%%%%%%%%%
Now, we use the master equation with atomic and cavity losses (for $n_{th}^{(at,cav)}=0$), as well as a phonon squeezed vacuum reservoir:
\begin{eqnarray}
    \frac{d\rho}{dt}&=&-\imath[\mathcal{\tilde{H}}_{3}+\mathcal{\tilde{H}}_{E},\rho]+\frac{\gamma_{21}}{2}\mathcal{L}[\sigma_{21}]\rho\nonumber\\
    &&
    +\frac{\gamma_{10}}{2}\mathcal{L}[\sigma_{10}]\rho+\frac{\kappa_{a}}{2}\mathcal{L}[a]\rho+\frac{\kappa_{b}}{2}\mathcal{L}_{sq}[b]\rho,
\end{eqnarray}
with the Lindbladian corresponding to the incoherent pump of squeezed phonons as $\mathcal{L}_{sq}[b]=(N_{sq}+1) \left(2b\rho b^{\dagger}- b^{\dagger}b\rho-\rho b^{\dagger}b\right) + N_{sq} \left(2b^{\dagger}\rho b- bb^{\dagger}\rho-\rho bb^{\dagger}\right)+M_{sq}\left(2b^{\dagger}\rho b^{\dagger}- b^{\dagger}b^{\dagger}\rho-\rho b^{\dagger}b^{\dagger}\right)+M_{sq}^{*}\left(2b\rho b-bb\rho-\rho bb\right)$; 
here $N_{sq}=\sinh^{2}{r}$ and $M_{sq}=-\exp{(\imath\theta)}\sinh{r}\cosh{r}$, obey the relation $\sqrt{N_{sq}(N_{sq}+1)}=|M_{sq}|$.

Using the above scheme, we explore the effects of the squeezing parameters ($r$, $\theta$) on the quantum fluctuations and fidelity, in order to achieve the best TSS. In the semiclassical approximation we get following results for the quadrature fluctuations in steady-state:
\begin{align}
    \left(\Delta x_{a}\right)^{2}&=\frac{1}{4}p+l,\\
    \left(\Delta x_{b}\right)^{2}&=\frac{1}{4}p\left[2\left(N_{sq}-M_{sq}\right)+1\right]+l,
\end{align}
where
\begin{align}
   p&=\frac{\kappa_{a}\kappa_{b}}{4J^{2}\langle\sigma_{21}^{\dagger}\rangle^{2}+\kappa_{a}\kappa_{b}},\\
   l&=\frac{J^{2}\langle\sigma_{21}^{\dagger}\rangle^{2}\left[\kappa_{b}\left(1+2\left[N_{sq}-M_{sq}\right]\right)+\kappa_{a}\right]}{\left[\kappa_{a}+\kappa_{b}\right]\left[4J^{2}\langle\sigma_{21}^{\dagger}\rangle^{2}+\kappa_{a}\kappa_{b}\right]}.
\end{align}
Now, we study the parameters in our hybrid system for which both quadratures reach their maximum squeezing and that allow the optimal TSS via optomechanical coupling. In panel $(a)$ of the Fig. \ref{fig4} the quadrature fluctuations are presented as functions of the parameter $r$. The results show that for $g_{cm}=0.01$, the TSS performed in an optimal way, since the curves are closer, that is, both fluctuations are similar. In panel $(b)$ of the Fig \ref{fig4}, we show the time evolution of the quadrature fluctuations. Without loss of generality, we have fixed the parameters $r=0.3$ and $g_{cm}=0.01$. Analogous to the aforementioned process of the coherent pump, we can see how the injection of a squeezed bath leads to cavity and mechanical squeezing in steady state.

\begin{figure}[t]
\centering
\includegraphics[width=0.494\linewidth]{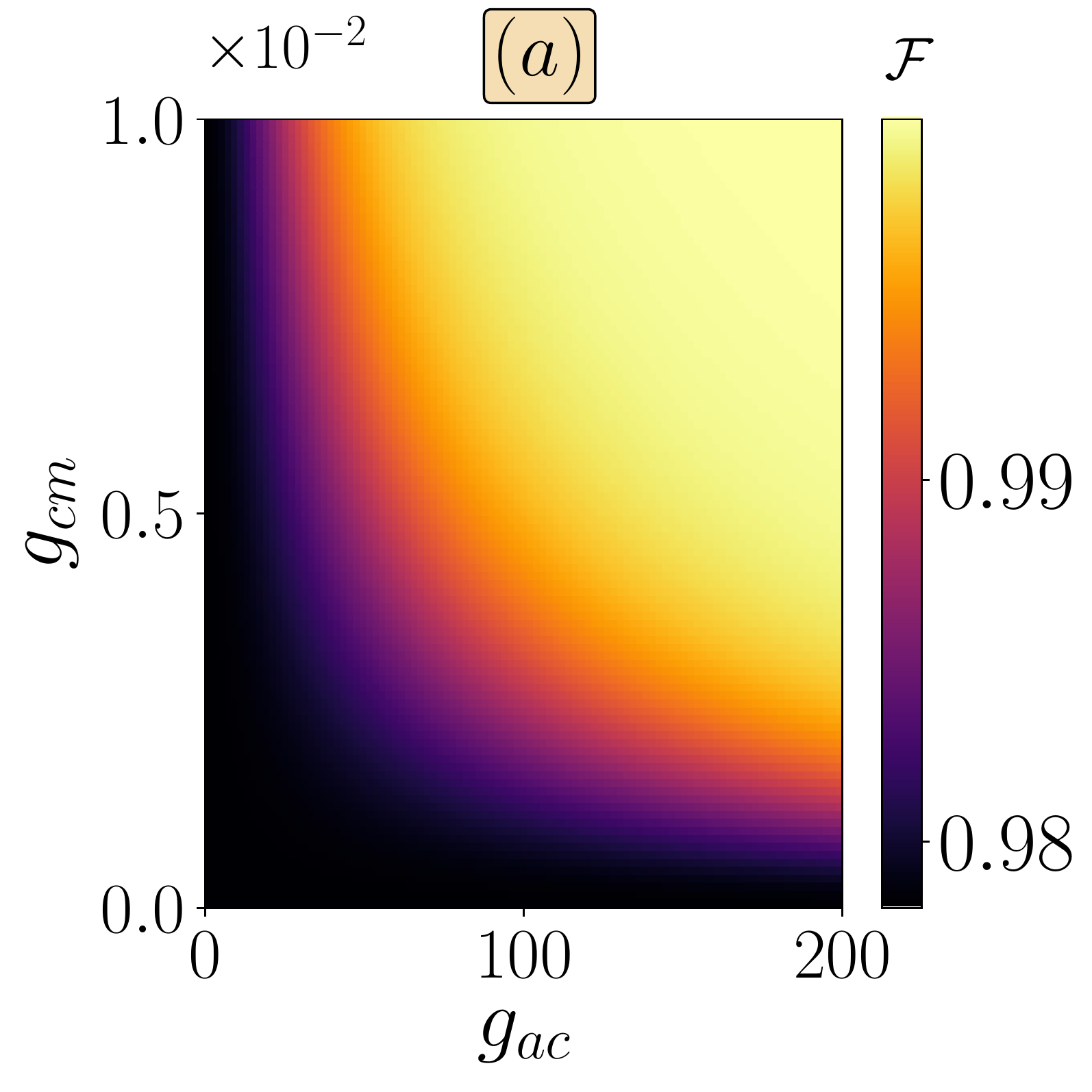}
\includegraphics[width=0.494\linewidth]{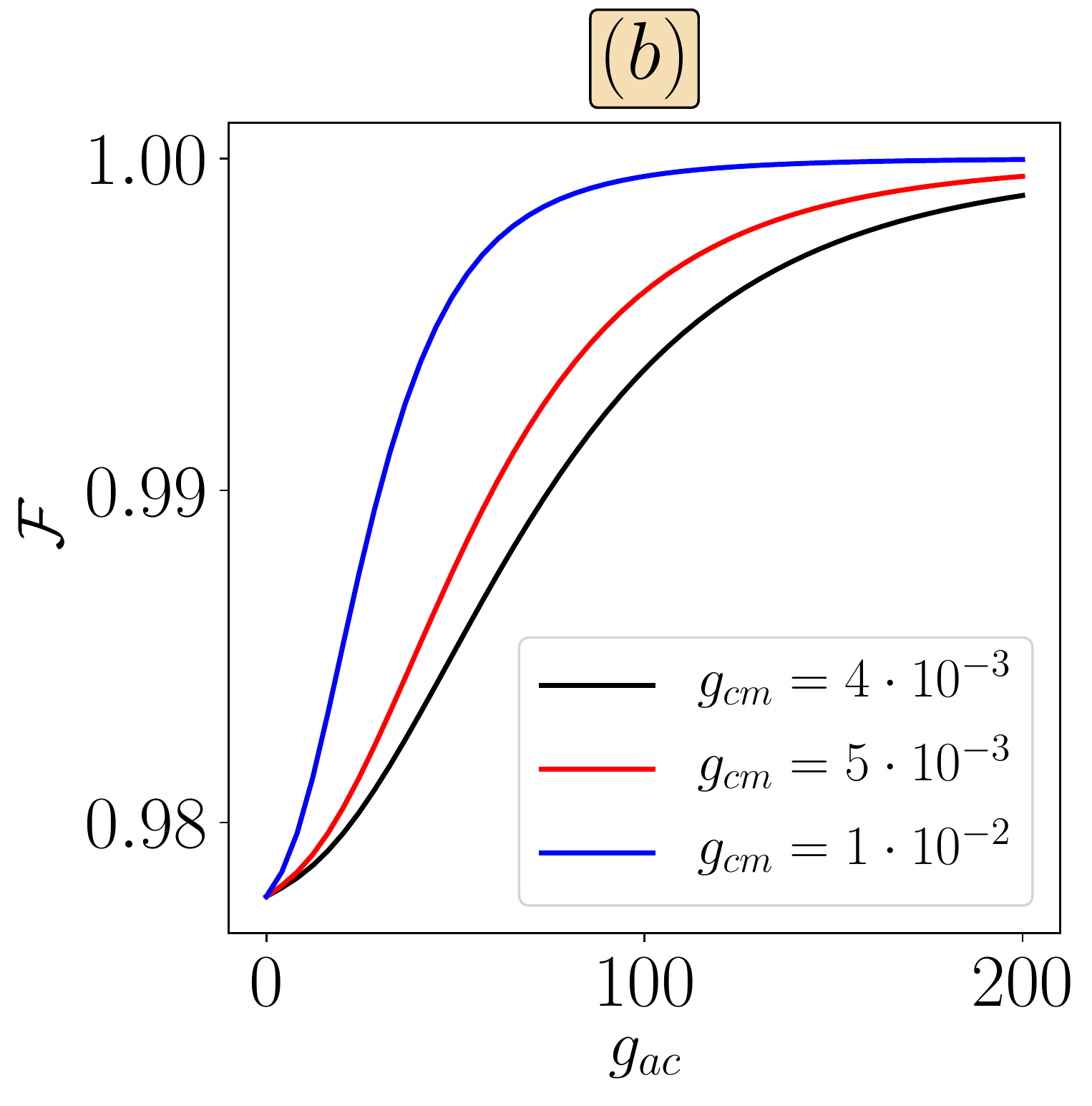}
\caption{$(a)$ Fidelity between the cavity field and mechanical state as a function of the couplings $g_{cm}$ and $g_{ac}$. $(b)$ Fidelity as a function of $g_{ac}$ for some couplings $g_{cm}$. As is observed, the fidelity is close to unity as the optomechanical and Jaynes-Cummings couplings increase. The parameters here are the same as in Fig. \ref{fig4}  with $r=0.3$.}
\label{fig5}
\end{figure}

In \cite{Wal2010} the authors propose a setup that allows the transmission from an atomic squeezed state to a membrane with a high fidelity in presence of relevant decoherence rates. Motivated for this, we simulate the influence of the Jaynes-Cummings coupling on the fidelity of the TSS in our system. In panel $(a)$ of Fig. \ref{fig5} we show the fidelity as a functions of $g_{cm}$ and $g_{ac}$. Notice that the fidelity is close to unity as the optomechanical and Jaynes-Cummings couplings increase, even in the presence of dissipation in the system. In panel $(b)$ of Fig. \ref{fig5} we show the fidelity as a function of $g_{ac}$ for some values of $g_{cm}$. This result allows us to conclude that in the strong coupling regime the open system allows the TSS with a reliability close to 100\%.

In conclusion, we have proposed a hybrid system consisting of a three-level atom, an optical cavity, and mechanical resonator that allows the steady-state squeezing conversion. We demonstrated that by pumping coherently and incoherently phonon squeezing in our hybrid system and also adding two coherent atomic drives, we can transfer steady-state squeezing to the cavity photons with an extremely high fidelity, close the unity (see Figs. \ref{fig3} and \ref{fig5}). 
%%%%%%%%%%%Figure 5%%%%%%%%%%%%%%
%%%%%%%%%%%%%%%%%%%%%%%%%%%%%%%
Furthermore, in the parameter range used in this work, as we increase the optomechanical coupling, we get better results in producing closer phonon and photon squeezed states, i.e. improving the protocol of TSS. The parameters used in this work are compatible with recent experiments for the optomechanical hybrid setups, e.g. \cite{RevModPhys.86.1391, Mir2020, Rie2016}. 
\begin{acknowledgments}
H.M. acknowledge financial support from Universidad Mayor through the Doctoral fellowship. V.E. and M.O. acknowledge the financial support from Fondecyt Regular No. 1180175.
\end{acknowledgments}

\bibliography{apssamp}% Produces the bibliography via BibTeX.

\end{document}